\documentclass[aps,pre,twocolumn,showpacs,showkeys,groupedaddress,floatfix]{revtex4-1}

\usepackage{amsmath}
\usepackage{amssymb}
\usepackage[utf8]{inputenc}
\usepackage{graphicx}
\usepackage{acronym}

\usepackage{soul} 
\usepackage{color}

\newcommand{\RR}{{\boldsymbol R}}

\begin{document}

\title{Structure and dynamics of the t154 lattice glass}

\author{Alejandro Seif}

\email{aseif@iflysib.unlp.edu.ar}

\author{Tom\'as S.~Grigera}

\affiliation{Instituto de F\'\i{}sica de L\'\i{}quidos y Sistemas
  Biol\'ogicos (IFLYSIB), CONICET and Facultad de Ciencias Exactas,
  Universidad Nacional de La Plata, Calle 59 no.\ 789, B1900BTE La
  Plata, Argentina}

\affiliation{CCT CONICET La Plata, Consejo Nacional de Investigaciones
  Cient\'\i{}ficas y T\'ecnicas, Argentina}

\affiliation{Departamento de Física, Universidad Nacional de La Plata}

\date{\today}
\def\xipts{\xi_\text{PTS}}

\begin{abstract}

  We revisit the $t154$ variant of the Biroli-Mezard lattice glass,
  complementing previous studies by studying statics and dynamics
  under periodic boundary conditions as well as systems confined in
  cavities with amorphous boundaries.  We compute the point-to-set
  correlation and relaxation times under the different boundary
  conditions.  Results point to a scenario with dynamics ruled by
  structural correlations.

\end{abstract}


\maketitle

\acrodef{KMC}{Kinetic Monte Carlo}
\acrodef{KCM}{kinetically constrained model}
\acrodef{TTI}{time-traslation invariance}
\acrodef{ABC}[ABCs]{amorphous boundary conditions}
\acrodef{PBC}[PBCs]{periodic boundary conditions}
\acrodef{BIC}{$\beta$ initial condition}
\acrodef{PTS}{point-to-set}
\acrodef{RFOT}{random first-order theory}
\acrodef{GC}{grand canonical}
\acrodef{C}{canonical}


\section{Introduction}

\label{sec:introduction}

The physical mechanism behind the dramatic slowing down of dynamics
close to the empirically defined glass transition has been subject of
continued interest and debate \cite{ediger_supercooled_1996,
  debenedetti_supercooled_2001, berthier_theoretical_2011,
  biroli_perspective:_2013}.  Due to the somewhat limited amount of
information available from experiments and simulations (limitations due in
large part to the difficulties that arise from the very phenomenon
under study, i.e.\ the slowdown), different theoretical
proposals have been able to rationalize observed behaviors often
starting from completely divergent viewpoints
\cite{chandler_dynamics_2010, gotze_relaxation_1992,
  lubchenko_theory_2007,biroli_random_2012}.  It is thus natural that
models have been sought that display the main phenomenology with a
bare minimum of ingredients, so as to allow for a more detailed
analysis (either theoretically or numerically), and lattice models
have been considered good candidates in this category
\cite{kob_kinetic_1993, biroli_lattice_2001,
  garrahan_geometrical_2002, dawson_dynamical_2003,
  ritort_glassy_2003, mccullagh_finite-energy_2005,
  pica_ciamarra_monodisperse_2003}.

Here we revisit a lattice glass model, the t154
\cite{darst_dynamical_2010}, a variation of the Biroli-M\'ezard
lattice glasses \cite{biroli_lattice_2001}.  Lattice glasses are
defined through an energy (which may be infinite) uniquely assigned to
every configuration, and glassy behavior follows from a ``natural''
dynamics (Metropolis Monte Carlo, for instance).  This is opposed to
kinetically constrained models \cite{ritort_glassy_2003}, where there
are no or few constraints to possible configurations, and glassy
dynamics results from rules that forbid certain transitions between
configurations.  A detailed study of dynamical heterogeneities of the
t154 was carried out in ref.~\onlinecite{darst_dynamical_2010}, where
it was found that it is stable against crystallization, and that it
has the main characteristics of a fragile liquid, showing in
particular Stokes-Einstein violations and signs of a growing
\emph{dynamic} length scale as measured by a four-point correlation
function.  In this respect, the t154 is phenomenologically similar to
kinetically constrained models.  Here we focus on an aspect left out
of this previous study, which is the determination of a \emph{static}
correlation length and its possible relationship with the dynamical
behavior.

We use the approach of studying small or confined systems to put the
relevant length scales in evidence.  To find a static (structural)
length scale we compute \ac{PTS} correlations
\cite{bouchaud_adam-gibbs-kirkpatrick-thirumalai-wolynes_2004,
  montanari_rigorous_2006}, which are computed by studying systems
confined in cavities with \ac{ABC} (explained below)
\cite{cavagna_numerical_2010}.  \ac{PTS} correlations were the first
to be used successfully to detect a growing correlation length in
supercooled liquids \cite{biroli_thermodynamic_2008,
  hocky_growing_2012, kob_non-monotonic_2012}, a result confirmed also
with other approaches \cite{widmer-cooper_irreversible_2008,
  karmakar_direct_2012, tanaka_critical-like_2010,
  sausset_characterizing_2011,Karmakar2014}.  For our dynamical
analysis we use both \ac{ABC} and the usual \ac{PBC}, looking for
changes in the relaxation times in small systems
\cite{berthier_finite-size_2012, Cavagna2012}.  The goal is to extend
the study of the t154 to its structural aspects, and to establish
whether the structural properties are relevant for the dynamical
features.

We present the model and details of our simulations in
sec.~\ref{sec:model-simulations}.  Sec.~\ref{sec:static-behavior}
summarizes our structural findings, sec.~\ref{sec:dynamics} is devoted
to our dynamical results, and we conclude in
sec.~\ref{sec:conclusions}.

\section{Model and simulations}

\label{sec:model-simulations}

The Biroli-M\'ezard lattice glasses \cite{biroli_lattice_2001} are
defined on a $d$-dimensional lattice. It's sites can be empty or
occupied by one and only one particle of class $\ell=1,2,\ldots$.  To
this hard excluded volume, a hard density constraint is added: at most
$\ell$ of the neighbouring sites are allowed to be occupied.
Different variants of the model arise when specifying the number of
classes and the proportion among them.  Here we focus in particular on
the t154 variant \cite{darst_dynamical_2010}, where $\ell=1,2,3$ and
the proportions of each class are 0.1, 0.5 and 0.4 respectively.
Since the constraints are hard, temperature is irrelevant and the
control parameter is the density $\rho$ or the Lagrange multiplier
$\alpha_1=\beta\mu_1$ of the particles of the first class ($\alpha_2$
and $\alpha_3$ being fixed by the composition).  The relationship
between chemical potential and composition can be written
\begin{equation}
  \label{eq:1}
  \beta \mu_\ell = \ln \left(\frac{\rho_\ell}{p_\ell}\right),
\end{equation}
where $p_\ell$ is the fraction $\ell$-holes, i.e.\ empty sites with
enough free neighbours that a particle of class $\ell$ can be placed
on it without violating the constraints.  Then, for fixed composition
one can determine $\mu_2$ and $\mu_3$ from the density and the
$p_\ell$ (obtained by simulation).

Dynamics slow down considerably at high densities (or for confined
systems as discussed below), so we have used Kinetic Monte Carlo
\cite{Gillespie1977,Bortz1975} as in \cite{darst_dynamical_2010} to
simulate the system, which brings a significant speed-up for all but
the lowest densities considered.  We have performed simulations both
in the canonical and \ac{GC} ensembles.  The GC ensemble allows us to
study systems more strongly confined than is possible with the
canonical ensemble, since in many cases cavities with ABCs get
completely stuck when simulating in the canonical ensemble.

\begin{figure}\includegraphics[width=\columnwidth]{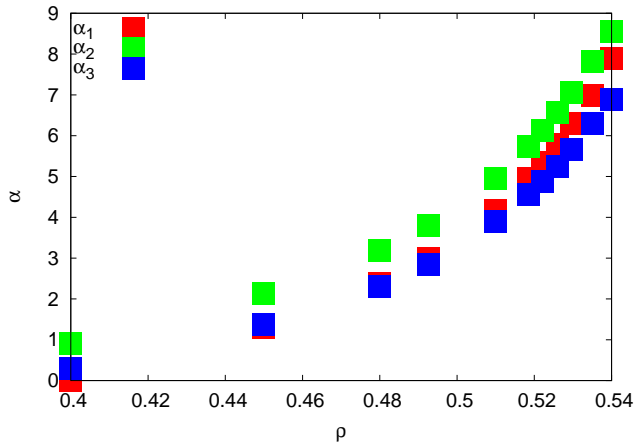}%
\caption{$\alpha$ vs $\rho$ obtained from canonical simulations @ $L=30$.}
\label{AlphaVsRho}
\end{figure}

When simulating in the \ac{GC} ensemble, care is required in choosing
the values of the chemical potential, especially at high densities, as
the composition is very sensitive to small changes in the
$\alpha_\ell$.  To determine these values we created valid
configurations of different sizes $L^3$ and densities $\rho$ with the
prescribed composition (by running a \ac{GC} simulation with very high
values of all the $\alpha_\ell$ and stopping as soon as the desired
density and composition were reached).  We then measured the number of
holes $p_\ell$ of each class (which is a natural output of the KMC
algorithm) in a canonical run and computed the $\alpha_\ell$ from
Eq.~\ref{eq:1} (see Fig.~\ref{AlphaVsRho}).  The composition of the
\ac{GC} runs was monitored to ensure it would not depart from the
desired proportion.

An important quantity in our analysis is the overlap $Q(t)$, defined
as
\begin{equation}
  Q(t)= \frac{1}{V_\mathcal{R}} \sum_{\RR \in \mathcal{R}} \langle
  n_\RR(t)n_\RR(0) \rangle,
  \label{overlapEQ}
\end{equation}
where $n_\RR$ is the occupation number of site $\RR$ ($n_R=0$ if empty
or $n_R=1$ if occupied by a particle of any class) and $V$ is the
volume (number of sites) of the region $\mathcal{R}$ included in the
sum.  $Q(t)$ is a measure of the correlation 
of the
region at time $t$ with itself at time 0 (with the time origin being
irrelevant in equilibrium).  The overlap of two configurations
independently drawn from a translation-invariant distribution is
$\rho^2$, and this is the value reached for $t\to\infty$ with
\ac{PBC}, indicating that correlation is lost.  For the region
$\mathcal{R}$ we take the whole lattice, the \ac{ABC} cavity (see
below), or a small cube in the center of the cavity (in which case we
name the overlap with a lowercase $q(t)$).

In \acl{ABC} the system is subject to a surface field applied at its
boundaries, which is created by particles of the same kind placed
outside the boundaries and held fixed in random positions drawn from
the equilibrium distribution.  In other words, one studies a cavity of
mobile particles surrounded by particles frozen at equilibrium
positions.  In practice this is achieved by taking an equilibrium
configuration obtained in a run with \ac{PBC} and artificially
freezing the particles outside a cubic cavity (but allowing the frozen
particles to interact with the mobile ones).  We use a system of size
$L^3=30^3$, in which we define a cubic cavity of size $K^3$.  By varying
the size $K$ one can study the effects of the boundary layer on the
statics and dynamics of the cavity.  Under these conditions the
asymptotic value of the overlap will not necessarily be the
uncorrelated value $\rho^2$.  The asymptotic value of the overlap at
the center $q_\infty\equiv \lim_{t\to\infty} q(t)$ is the
\emph{point-to-set correlation,} and is a measure of the influence of
the boundary of the cavity on the structure at its center.  To measure
the PTS correlation we used a cube of side 3, and averaged over 50
realizations of the boundaries.  When reporting ABC results, the
global overlap $Q(t)$ is computed only within the cavity (mobile
particles).

To ensure that our runs are long enough that the asymptotic value of
the PTS, $q_\infty$, represents the equilibrium value and is not the
result of running the simulation for too short times, we perform a
$\beta$-initial condition (BIC) test \cite{Cavagna2012}.  For this we
initialize two identical cavity samples in a configuration $\gamma$
that will serve as the reference against which the instantaneous
overlap is computed.  In one configuration the cavity particles are
replaced by those of a different configuration $\delta$ with very low
overlap with $\gamma$.  If $q(t)$ reaches the equilibrium value we
should see $q_{\gamma\gamma}(t)$ decrease toward $q_\infty$, while
$q_{\gamma \delta}(t)$ will increase up to $q_\infty$.  If the two
samples do not reach the same $q_\infty$, thermalization of the sample
has not been achieved (negative BIC test).  In practice, it is easy to
do the test in the GC case by simply emptying the cavity after taking
the initial configuration as reference.


\section{Results}
\label{sec:results}

\begin{figure}\includegraphics[width=\columnwidth]{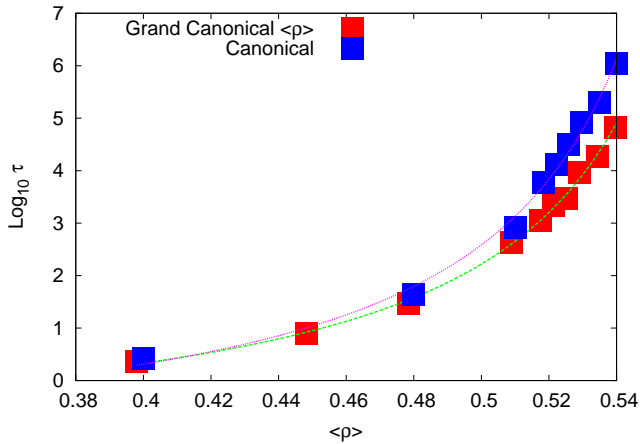}
  \caption{VFT fit for BULK C/GC dynamics. L=30. For
    $\rho \geq 0.51$ the difference between dynamics becomes broader.}
  \label{AngellPlot-BULK-L30}
\end{figure}

Our aim is to establish possible connections between the spatial
structure and the dynamical behavior, so we measure space and time
correlations of the density, as encoded in the overlap
(Eq.~\ref{overlapEQ}).  We start showing a (generalized
\cite{berthier_compressing_2009}) Arrhenius plot of the relaxation
time vs.\ the density for both \acf{C} and \acf{GC} dynamics
(Fig.~\ref{AngellPlot-BULK-L30}).  The relaxation times $\tau$ were
extracted from a stretched-exponential fit of the time decay of the
local overlap,
\begin{equation}
  \label{eq:2}
   q_c(t)= A \exp[-(t/\tau)^{\beta}] + q_\infty.
\end{equation}
The plot reveals the fragile character of the model, and the curves
can be fitted by (generalized) Vogel-Fulchner-Tamman function
$\tau=\exp[A/(\rho-\rho_K)]$, yielding $\rho_K ^{GC}= 0.595$ and
$\rho_K^{C}=0.588$.

Since the model is defined with hard constraints (as the hard spheres
model for instance), at high densities the dynamics will start to
become sluggish because to relax a configuration the system must find
a path that goes through allowed configurations (otherwise the energy
price is infinite). However, these configurations are becoming less numerous because the
constraints are harder to fulfill the higher the density or the
stronger the confinement.  At still higher densities, groups of
configurations can become completely disconnected (i.e.\ separated by
infinite-energy barriers), and the system becomes nonergodic.  On the
other hand, the (nonphysical) dynamics of the \ac{GC} ensemble allows
destruction and creation of particles at arbitrary locations, thus
effectively lowering barriers by adding connections between
configurations. In particular, the loss of ergodicity is avoided,
because in the worst case scenario two configurations could be joined
by a path that first destroys all particles and then creates them in
the required locations.  Thus one expects shorter relaxation times
with respect to the canonical dynamics, at least at relatively high
densities where the canonical dynamics start slowing down because
many trial moves lead to forbidden configurations.  This expectation
is fulfilled, but the \ac{GC} times are appreciably smaller only for
$\rho\gtrsim \rho_0 \approx 0.51$.  We take this as an indication
that the structure has important influence on the dynamics only for
densities greater than $\rho_0$, which would indicate the start of
``landscape influenced'' dynamics.

\subsection{Structure}

\label{sec:static-behavior}

\begin{figure}\includegraphics[width=\columnwidth]{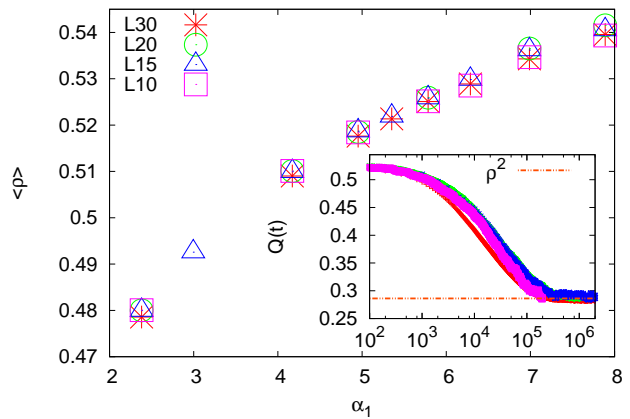}%
  \caption{Density $\rho$ vs $\alpha_1$ in PBC GC dynamics. The higher
    $\alpha_l$ produce $\rho$ that vary with $L$. (The $L=30$ data are
    the same shown in Fig.~\ref{AlphaVsRho}). Inset: Overlap
    $Q(t)$ vs time for different box sizes $L$ in PBC GC dynamics
    ($\alpha$ constant). As expected, $Q(t)$ for PBC GC decays to
    $\rho^2$ when thermalized.}
\label{Alpha-Density-PBC-GC}
\end{figure}

\begin{figure}\includegraphics[width=\columnwidth]{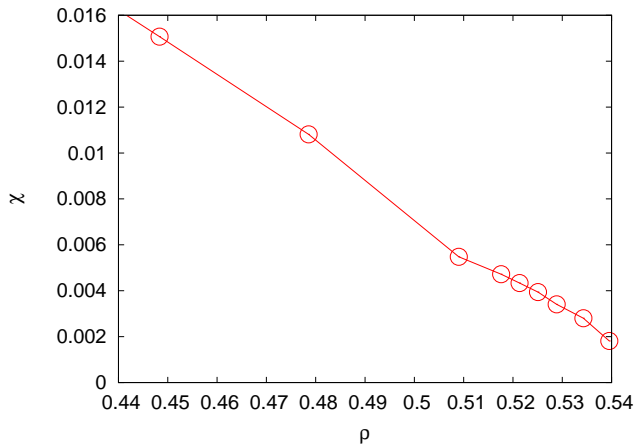}%
  \caption{Susceptibility $\chi$ vs $\langle \rho \rangle$ for PBC
    GC. System displays a decreasing $\chi$ througout all $\rho$ values.}
\label{Susceptibility-PBC-GC}
\end{figure}

We start by plotting the density $\rho$ vs.\ the Lagrange multiplier
$\alpha_1=\beta \mu_1$ (the logarithm of the fugacity of particles of
class 1), Fig.~\ref{Alpha-Density-PBC-GC}.  We find slight dependence
on size for $\alpha_1>4$.  However, the dependence is not monotonic
with $L$ as is typical of finite-size effects.  We also computed (through
fluctuations) the susceptibility
\begin{equation}
  \label{eq:3}
  \chi_\ell = \frac{\langle N_\ell ^2\rangle -\langle N_\ell
    \rangle^2}{V},
\end{equation}
where $\ell$ indicates particle class and we use $\chi$ without
subscript for the susceptibility corresponding to the total number of
particles (Fig.~\ref{Susceptibility-PBC-GC}).  These quantities show
no sign of singular behavior near $\rho_0$.  In particular, there is
no sign of a growing length scale: since $\chi$ is the volume integral
of the connected density correlation,
$C_c(r) = \langle \rho(0)\rho(r)\rangle - \langle\rho\rangle^2$, a
growing correlation length would cause an increase of the normalized
integral $\int\!\!dV\, C_c(r)/C_c(0) = V \chi/\langle N^2\rangle$.

This absence of order as detected by two-point correlation functions
while the relaxation times grows is typical of supercooled liquids.
For such systems, it has been shown that it is the \acf{PTS}
correlation that can detect the presence of order.  This is an
``agnostic'' measure of order, in the sense it does not make
assumptions about the order parameter, or about the kind of order that
is developing.

\begin{figure}\includegraphics[width=\columnwidth]{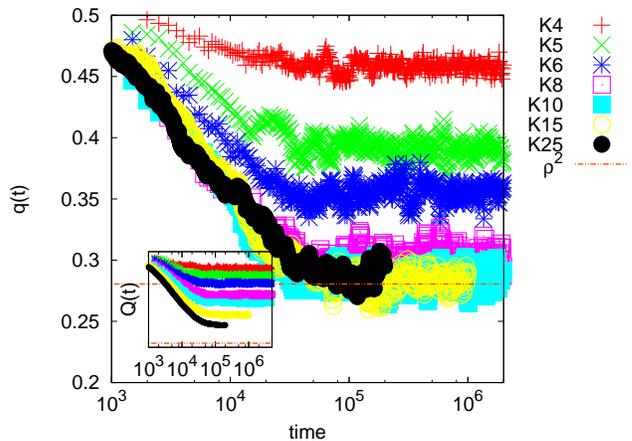}%
  \caption{Large: q(t) and Inset: Q(t) vs time for $\langle \rho \rangle=0.526$ for
    cavities of size $K^3$.  In the Overlap, all asymptotic values
    fall higher than bulk value (given by $\rho^2$), whereas in the
    PTS only the smalles cavities stay above it.}
\label{multiplotPTSKOVERLAPvst}
\end{figure}

To find the \ac{PTS} we computed the decay of the self-overlap
(Eq.~\ref{overlapEQ}) for systems at different chemical potentials and
confined in cubic cavities of side $K$ with \ac{ABC}.  Both the global
overlap $Q(t)$ (the overlap of the full cavity with itself) and the
overlap $q(t)$ of a small cube at the center of the cavity were
computed (see Fig.~\ref{multiplotPTSKOVERLAPvst}).  The \ac{PTS}
correlation is obtained as the $t\to\infty$ limit of $q(t)$.  All the
results of this section were obtained in the \ac{GC} ensemble, where
the dynamics are faster and allows us to equilibrate systems with
densities up to $\rho = 0.54$.  Dynamics in the canonical ensemble are
too slow and it is impossible to equilibrate even moderately confined
cavities in canonical runs (see sec.~\ref{sec:dynamics}).  However,
the choice of dynamics is irrelevant for the structural results
(provided the system can be thermalized).  Given that we have found
some size dependence of the chemical potential, we have checked that
the composition of the cavities stays at the 1-5-4 proportion.  We
have found some fluctuation in the composition of the smallest
cavities, but in no case larger than 3\%.

\begin{figure}\includegraphics[width=\columnwidth]{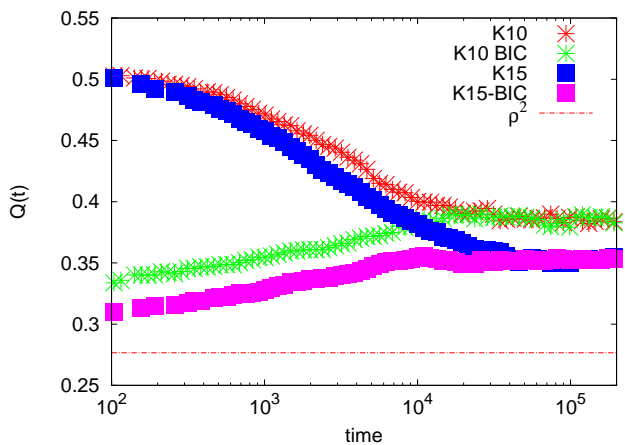}%
  \caption{BIC test for $\rho=0.526$ for cavities of size $K^3=10^3$ and
    $K^3=15^3$. We can see how the test renders a positive result, since
    both reach the same asymptotic values.  Therefore we conclude that
    both configurations have thermalized.}
  \label{BIC}
\end{figure}

When the structure can decorrelate completely (as when thermalized
under \ac{PBC}), we know the asymptotic limits of the overlaps
($\langle \rho\rangle^2$ for both $q(t)$ and $Q(t)$, see
Fig.~\ref{Alpha-Density-PBC-GC}, inset.).  The presence of structural
correlations is revealed by the fact that the asymptotic value for the
cavity is higher than the \ac{PBC} case.  Of course, a simulation that
is too short to thermalize the system could produce a spurious high
value of the asymptotic overlap.  To check that the system has indeed
equilibrated (and that we are measuring the actual equilibrium
\ac{PTS}), we perform \ac{BIC} tests \cite{cavagna_mosaic_2007} (see
sec.~\ref{sec:model-simulations}).  One such test is shown in
Fig.~\ref{BIC}: the same value of the overlap is reached starting from
independent (low overlap) configurations.  We then compute the
asymptotic values $q_\infty$ and $Q_\infty$ averaging the overlap
starting from the time when the two curves of the \ac{BIC} test
coincide.

From both $Q(t)$ and $q(t)$ it is clear that in small cavities the
border is exerting a significant influence on the particles inside.
We will use the values $q_\infty(K)$ and $Q_\infty(K)$ to extract a
correlation length.



\subsection{Correlation length}
\label{sec:correlation-length}

\begin{figure}\includegraphics[width=\columnwidth]{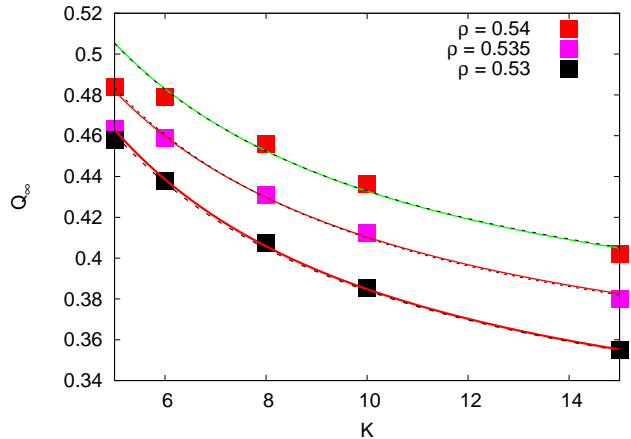}%
  \caption{Measurement of PTS via $q_{\infty}$ vs.\ cavity size
    $K$.  Full lines: fit to Eq.~\ref{q_c}.  Dashed lines:fit with Eq.~\ref{qtot}. }
  \label{Qtot-DoubleFit}
\end{figure}

Both $Q_\infty(K)$ and $q_\infty(K)$ contain information about the
structural correlations, but although the former is less noisy, it is
more difficult to interpret.  We start with the \ac{PTS} correlation
$q_\infty(K)$ (Fig.~\ref{Qtot-DoubleFit}).  We find that a simple
exponential can adequately fit the decay of the \ac{PTS}, from which
we extract a correlation length $\xipts$:
\begin{equation}
q_c(K)-q_0=(q_1-q_0)\exp[-K/ \xipts],
 \label{q_c}
\end{equation}
where $q_0=\langle\rho\rangle^2$ and $q_1$ and $\xipts$ are fit
parameters.  When $K\gg\xipts$, the center of the cavity is free to
rearrange as if it were subject to \ac{PBC}, thus $\xipts$ measures
how far the local structure influences the arrangement of other
particles.  $\xipts$ is found to increase steeply for $\rho>0.53$.
However, we find no sign of the nonexponential behavior described in
refs.~\onlinecite{biroli_thermodynamic_2008, hocky_growing_2012,
  gradenigo_static_2013}, explained invoking the appearance of the
multiple metastable states of the Random First-Order theory of liquids. In that picture, large cavities can explore all metastable states as
the liquid does, but small cavities are locked in to one state by
surface tension
\cite{bouchaud_adam-gibbs-kirkpatrick-thirumalai-wolynes_2004}.

\begin{figure}\includegraphics[width=\columnwidth]{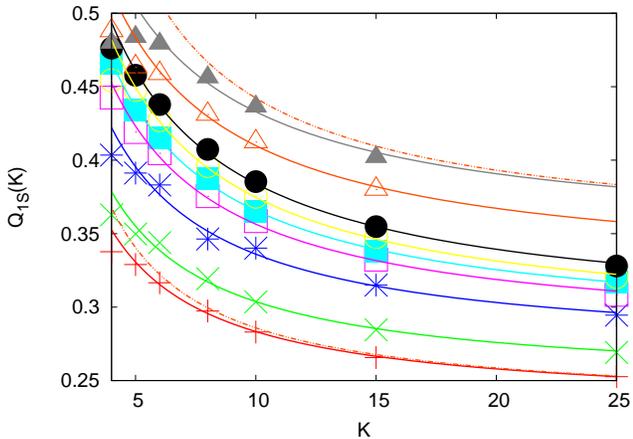}%
  \caption{From bottom to top, 
  static $Q_\infty(K)$ for $\rho=0.48,0.493,0.51,0.518,0.522,
  0.526,0.5296,0.535,0.54$. Fitted using $Q_{1S}(K)$ from eq.~\ref{Q1S}. 
  System displays 1-State behaviour for $K>5$. 
  Dashed lines are the asymptotic $1/K$ behavior of Eq.~\ref{Q1S} 
  for $\rho=0.48,0.535$.}
  \label{KOverlap-Asymptotic}
\end{figure}

\begin{figure}
  \includegraphics[width=\columnwidth]{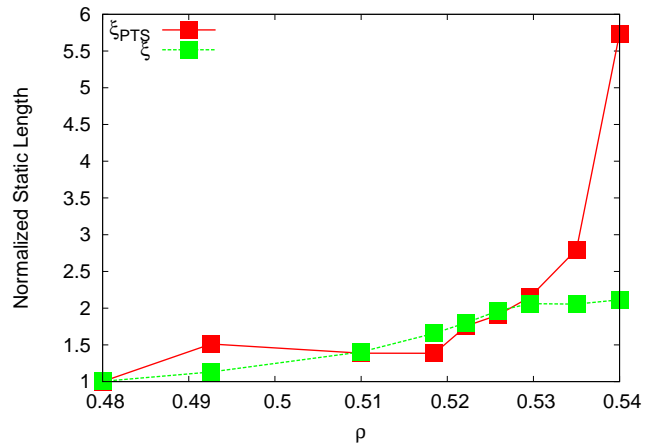}%
  \caption{Normalized Correlation length $\xi_S$ for the Overlap and
    PTS relaxation.  Both display a smooth increase to duplicate their
    low density value, however, the PTS length displays a sudden
    growth for $\rho>0.53$}
\label{LongCorrNormalized}
\end{figure}

The structural information contained in $Q_\infty(K)$ is encoded in a
more complicated way: even for cavities several times larger than
$\xipts$ (i.e.\ for distances over which there is no correlation as
measured by the \ac{PTS}), $Q_\infty$ will be higher than the \ac{PBC}
value due to the pinning effect of the frozen border on particles near
the edge of the cavity.  That is, even when the center of the cavity
is completely uncorrelated with the border, the global overlap is
picking up the influence of the border over the nearby particles.
If one makes the simple ``one state'' assumption (i.e.\ there are no
metastable states such that the cavity is always in the only liquid
state, and the overlap decays exponentially from a value of 1 at the
border to a value $q_0$ well inside the cavity) one gets
\cite{cavagna_mosaic_2007}
\begin{equation}
  Q_{1S}(K)=3(1-q_0)\left[ \frac{1}{x} - \frac{2}{x^2} +
    \frac{2(1-e^{-x})}{x^3}\right] + q_0,
\label{Q1S}
\end{equation}
with $x=K/\xi$, where the penetration length $\xi$ should be
proportional to $\xipts$ in this scenario.  A fit of this expression,
shown in Fig.~\ref{KOverlap-Asymptotic}, yields a penetration length
that behaves as shown in Fig.~\ref{LongCorrNormalized}.  Comparing the
evolution of $\xipts$ and $\xi$ with density, both show qualitatively
similar behavior up to $\rho\approx0.53$, increasing approximately
two-fold from $\rho=0.48$.  At higher densities their behavior differs
markedly: while $\xi$ stays constant, $\xipts$ increases steeply
(approximately three times from $\rho=0.53$ to $\rho=0.54$).

\subsection{Dynamics}

\label{sec:dynamics}

We now compute the relaxation times $\tau$ under various conditions.
In all cases it was extracted from a stretched-exponential fit
($q_c(t)= A \exp(-(t/\tau)^{\beta} + q_\infty$) of the time decay of
the local overlap $q_c(t)$.  The aim is to study size effects on
$\tau$, to see whether a dynamically relevant length scale can be
detected.  As opposed to the statics study of
sec.~\ref{sec:static-behavior}, here the choice of canonical or
grand-canonical ensemble can make a big difference.  The hard
constraints of the model are hard to fulfill the higher the density or
the stronger the confinement, thus we expect confined systems with canonical
dynamics will have larger relaxation times as the cavity is made
smaller, and eventually become completely jammed
\cite{barnett-jones_transition_2013}.  On the other hand, the
(nonphysical) dynamics of the \ac{GC} ensemble lowers barriers and
avoids ergodicity breaking as discussed above.  Thus one expects
shorter confined systems with \ac{GC} dynamics to be faster, but also
the shape of the relaxation time vs.\ size curve could be
qualitatively different.

\subsubsection{GC dynamics}

\begin{figure}
  \includegraphics[width=\columnwidth]{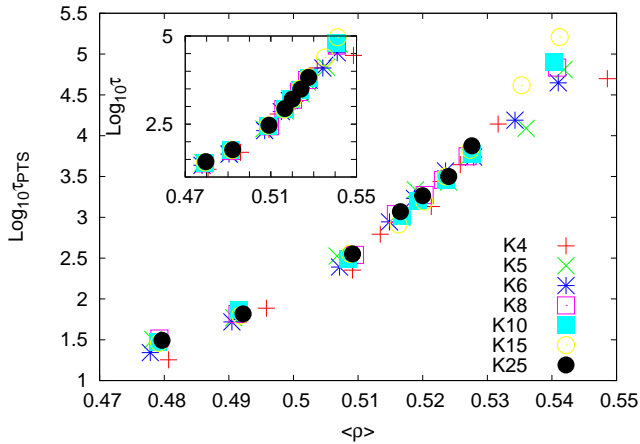}%
  \caption{PTS Relaxation time $\tau$ vs $\langle \rho \rangle$ using ABC GC. For $\langle
  \rho \rangle \geq 0.53$, $\tau$ begins to scatter by changing cavity size $K$. INSET: Overlap Relaxation time $\tau$ vs $\langle \rho \rangle$ using ABC GC.}
  \label{TauDensityGCPTS}
\end{figure}

From the decay of $q(t)$ for \ac{GC} dynamics in \ac{ABC} cavities we
obtained the relaxation times shown in Fig.~\ref{TauDensityGCPTS},
with a stretching exponent $\beta$ taking values in $0.45 < \beta < 0.85$ 
for the smaller cavities ($K \leq 10$), whereas for larger cavities ($K > 10$) 
$\beta$ smoothly shifted to the $0.50 < \beta < 0.65$ region. 
There are no discernible size effects up to
$\rho\approx 0.53$ (the density at wich $\xipts$ starts to grow
rapidly).  Beyond this density, scatter among the curves is seen, with
smaller cavities seemingly faster than larger ones.  Unfortunately
there are small variations in $\rho$ as $K$ is varied, which
forbids from plotting $\tau$ vs $K$ at constant density (because
although the fluctuations in $\rho$ are small $\tau$ is extremely
sensitive to $\rho$).


\begin{figure}\includegraphics[width=\columnwidth]{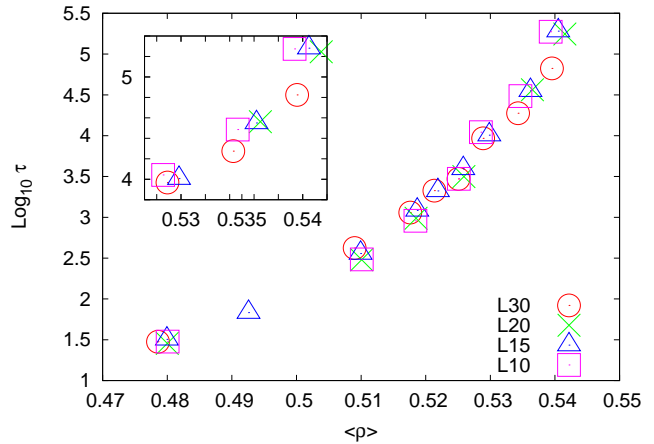}%
  \caption{Relaxation time $\tau$ vs $\langle \rho \rangle$ using PBC
    GC. Larger systems appear to relax faster in high density
    configurations. INSET: Zoomed region shows density variations for
    each box size L. }
\label{TauDensityGC-PBC}
\end{figure}

We did the same analysis for \ac{PBC} (Fig.~\ref{TauDensityGC-PBC}).
In this case, although there is no clear tendency and density seems to 
be more scattered than in \ac{ABC}, larger systems appear to relax faster.

\subsubsection{Canonical dynamics}
\label{sec:canonical-dynamics}

Turning now to canonical dynamics, as discussed above one expects
smaller systems to be slower and eventually completely jammed.  Indeed we
found that as size is decreased, it is more and more likely to find a
sample that is stuck out of equilibrium, i.e., the value of $q(t)$ or
$Q(t)$ oscillates at values higher than the equilibrium value found
with \ac{GC} dynamics and validated with the \ac{BIC} test.  Sometimes
this value is nearly $Q(t)=\rho$, i.e.\ the $t=0$ value.  The results
we report here were obtained by averaging only over samples that
\emph{do not} block, following the logic of
ref.~\onlinecite{berthier_finite-size_2012}.  The idea is that this is
a toy model with hard core interactions, which therefore artificially
excludes relaxation mechanisms (like activated jumps) which a real
system could use in order to relax.  Thus the jammed samples would
contribute to the average an exaggerated (infinite) relaxation time,
so the average obtained excluding them would give a trend
qualitatively more similar to the behavior of a realistic model.  We
report only \ac{PBC} results for $L\ge10$, since systems with $L<10$
systems tend to block, e.g., for $\rho=0.526$, only two out of fifty
samples would relax for $L=6$.  We do not report systems confined with
\ac{ABC} because in those cases the relaxation time grows very quickly
and most samples end up completely jammed, even for the largest
cavities.

\begin{figure}\includegraphics[width=\columnwidth]{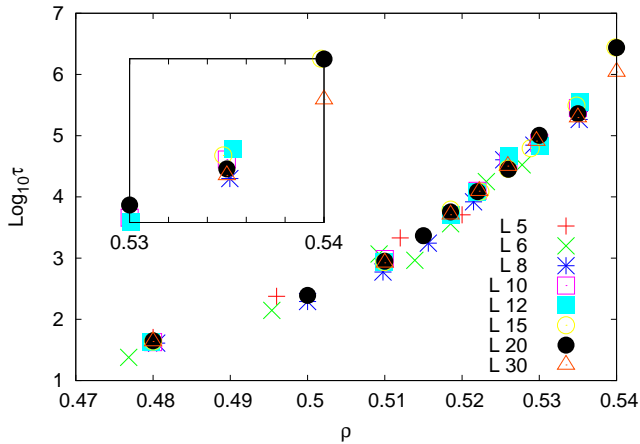}%
  \caption{Relaxation time $\tau$ vs $\rho$ using Canonical PBC for
    various boxes of size $L^3$, displays larger relaxation times than
    GC. INSET: Zoomed region shows subtle differences in density
    between $L$.}
\label{TauDensityC}
\end{figure}

\begin{figure}\includegraphics[width=\columnwidth]{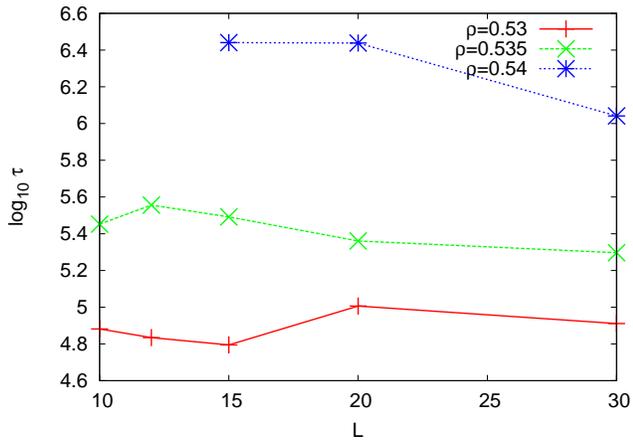}%
  \caption{Relaxation time $\tau$ vs $L$ in PBC Canonical dynamics,
    for $\rho \simeq 0.53,0.535,0.54$}
\label{CanoTauvsL}
\end{figure}

Fig.~\ref{TauDensityC} shows $\tau$ vs.\ $\rho$ for \ac{PBC} systems
of different sizes.  For $\rho<0.53$ there are no size effects, except
perhaps for the smallest ($L=5$) system.  At the two highest
densities, it seems the smaller systems are \emph{slower} than larger
ones (i.e.\ the opposite from the trend observed in \ac{ABC} cavities
with \ac{GC} dynamics).  However, a more complicated (nonmonotonic)
behavior is to be expected with canonical dynamics
\cite{berthier_finite-size_2012, Cavagna2012}.  Some evidence of this
is Fig.~\ref{CanoTauvsL}.

\section{Discussion and conclusions}
\label{sec:conclusions}

We have presented a study of some dynamic and thermodynamic properties
of the t154 model that complements the analysis of
ref.~\cite{darst_dynamical_2010}.  At the static level, we have
computed the \acf{PTS} correlation, and found it decays as a simple
exponential with the size of the cavity.  The corresponding
correlation length $\xipts$ shows around $\rho\approx0.53$ a rather
steep increase (about threefold between $\rho=0.525$ and $\rho=0.54$).
Up to $\rho=0.53$, the trend in \ac{PTS} correlation length $\xipts$
is the same as that of the lengthscale $\xi$ obtained by fitting the
global overlap of the cavity $Q(K)$, assuming a simple exponential
decay of the local overlap from the wall boundary inside.  Above this
density, $\xi$ continues to rise gently and does not reflect the steep
increase of $\xipts$.

The sharp increase of $\xipts$ together with the breakdown of
proportionality between $\xipts$ and $\xi$ seems to point to
$\rho\approx 0.53$ as a density marking a change in behavior, perhaps
a breakdown of the one-state scenario.  Thus for
$\rho= 0.53, 0.535, 0.54$ we have tried to fit $Q(t)$ with an
alternative to the one-state formula~\eqref{Q1S}.  In a multistate
(\ac{RFOT}) scenario, one can still assume that the overlap will decay
exponentially, but it can reach the value $q_0$ or some other (higher)
value $q_1$, depending on whether the cavity is free or locked into
one state.  Combining the two possibilities with their Boltzmann
weight \cite{biroli_thermodynamic_2008} gives:
\begin{equation}
 Q_\text{MS}=Q_0(K)+[Q_1(K)-Q_0(K)] \exp{\left[ -K/\xipts \right]} ,
 \label{qtot}
\end{equation}
where the $Q_n(K)$ have the form of Eq.~\eqref{Q1S} but with different
asymptotic values $q_0$ and $q_1$, and possibly two different
penetration lengths $\xi_{1,2}$.  We used Eq.~\eqref{qtot} to fit
the $Q_\infty(K)$ for the three highest densities, taking $\xipts$
from the $q_\infty(K)$ fit, fixing $q_0=\langle\rho\rangle^2$ and
$\xi_1=\xi_2=\xi$.  This two-parameter fit is good but yields $q_1\approx0.3 \approx q_0$, i.e.\ it essentially recovers the one-state fit.  This together with the purely
exponential relaxation of $q_0(R)$ makes it hard to invoke a multistate scenario.

We have also performed a finite-size study of the relaxation times (as
obtained from the decay of the self overlap) using both \ac{GC}
dynamics (where we examined systems with \ac{ABC} and \ac{PBC}) and
canonical dynamics (where only \ac{PBC} systems could be studied).
The \ac{GC} dynamics start showing finite-size effects (or
fluctuations at least) around the density (0.53) at which $\xipts$
starts growing steeply.  This is in agreement with a structural, rather
than kinetic origin of the slowdown mechanism of this model.


The analysis of canonical dynamics also shows some indication of
finite-size effects around density $\rho = 0.53$, but this analysis is
rather inconclusive, mainly because it has not been possible to reach
high densities with small systems.  The trend is apparently the
opposite with respect to \ac{GC} (i.e.\ smaller systems are slower
rather than faster).  However very small systems should eventually
become faster \cite{berthier_finite-size_2012}, a nomonotonicity which
we have not clearly observed.

In summary, our results hint at a scenario where dynamics are ruled by
structural correlations, but with little evidence for a particular
theory.


\begin{acknowledgments}

We thank G.~Parisi for discussions on the t154 model.  This work was
supported by grants from Consejo Nacional de Investigaciones
Científicas y Técnicas (CONICET, Argentina), Agencia Nacional de
Promoción Científica y Tecnológica (ANPCyT, Argentina), and
Universidad Nacional de La Plata.

\end{acknowledgments}

\bibliographystyle{apsrev4-1}
\bibliography{main.bib}

\end{document}